\documentclass[twocolumn]{aastex6}

\shorttitle{Solution for Orbiting Wind}
\shortauthors{Wilkin \& Hausner}

\begin{document}


\title{Exact Analytic Solution for a Ballistic Orbiting Wind}


\author{Francis P. Wilkin and Harry Hausner\altaffilmark{1}}
\affil{Union College\\
Schenectady, NY 12308, USA}





\altaffiltext{1}{Present address: Department of Physics, University of Wisconsin-Madison}

\begin{abstract}
Much theoretical and observational work has been done on stellar winds within binary systems. We present 
a new solution for a ballistic wind launched from a source in a circular orbit. Our method emphasizes the 
curved streamlines in the corotating frame, where the flow is steady-state, 
allowing us to obtain an exact solution for the mass density at 
all pre-shock locations. Assuming an initially isotropic wind, 
fluid elements launched from the interior hemisphere of the wind will be the first to cross other streamlines, 
resulting in a spiral structure bounded by two shock surfaces.  Streamlines from the outer wind hemisphere later intersect 
these shocks as well.  An analytic solution is obtained for the geometry of the two shock surfaces. 
Although the inner and outer shock surfaces asymptotically trace Archimedean spirals, 
our tail solution suggests many crossings where the shocks overlap, beyond which the analytic solution cannot be continued. 
Our solution can be readily extended to an initially anisotropic wind.

\end{abstract}




\section{Introduction} \label{sec:intro}
Virtually all stars have a wind, and a large fraction of stars are 
in binary or multiple systems \citep{2013ARA&A..51..269D}, 
where the orbital motion will influence the flow of the wind. 
The motion of the driving source results in a ``reflex'' 
perturbation somewhat similar to parallax, whereby the flow is  altered  due to the changing 
location and velocity of the source \citep{Soker94,Canto99,He07,Raga11}.


Binarity has been invoked as the cause of a wealth of structures in planetary nebulae and their 
asymptotic giant branch (AGB) precursors \citep{M81,L82,Soker94,MM99}. Dramatic examples include 
spiral nebulae such as AFGL 3068 \citep{MauronHuggins,Morris2006} and CIT 6 \citep{Kim13}.  
Detailed analyses have been presented to use the nebular 
morphology and kinematics to diagnose unresolved features of the drivers, both to infer binarity and 
binary separation, as well as to determine wind properties \citep{KimTaam2012c,Homan15}. 
The gravitational wake mechanism \citep{Kim2011,KimTaam2012a}
can also produce a spiral pattern, but has a more flattened morphology, allowing one to in principle 
distinguish the presence of two separate, spiral morphologies. 
Additionally, the  presence of two winds in a binary system leads to a collision region bounded by two shocks, 
which may develop mixing in the presence of a wealth of instabilities 
\citep{MZ91,Monnier99,Lamberts12,RN16,Hendrix16}. 

We present a new, exact solution to the problem of an
initially isotropic wind launched ballistically from an orbiting point source in a circular orbit. 
In \S\ref{sec:formulation} we describe the streamlines and kinematics, in \S\ref{sec:density} we derive 
the density structure, and \S\ref{sec:discuss} discusses the shock criterion and the 
implications of this work.

\section{Mathematical Formulation} \label{sec:formulation}

We consider a wind launched from an orbiting star. 
Although the motion will clearly be time-dependent, a steady-state flow pattern 
should exist in the reference frame corotating with the star's orbital motion about the center of mass, 
provided the orbit is circular and the wind launch conditions are unchanging in that frame.  
Let the star orbit a distance $R_\circ$ from the origin in the X-Y plane, at angular 
frequency $\omega$ and counterclockwise as viewed from above.  
In the inertial frame the star's position and velocity as a function of time are 
\begin{eqnarray}
{\bf R}_*(t) & = & R_\circ(\hat{\bf x} \cos\omega t + \hat{\bf y} \sin\omega t),\label{eq:rstar} \\
{\bf V}_*(t) & = & \omega R_\circ (-\hat{\bf x} \sin\omega t + \hat{\bf y} \cos\omega t).\label{eq:vstar}
\end{eqnarray}
Here $\hat{\bf x}$ and  $\hat{\bf y}$ are the usual unit vectors along the $X$ and $Y$ axes. 
The Cartesian coordinates in the inertial and corotating (primed) frames are related by $Z=Z'$ and 
\begin{eqnarray}
X & = & X' \cos\omega t - Y' \sin\omega t,\nonumber\\
Y & = & X' \sin\omega t + Y' \cos\omega t.
\end{eqnarray}

\begin{figure}[ht!]
\includegraphics[width=8.5cm]{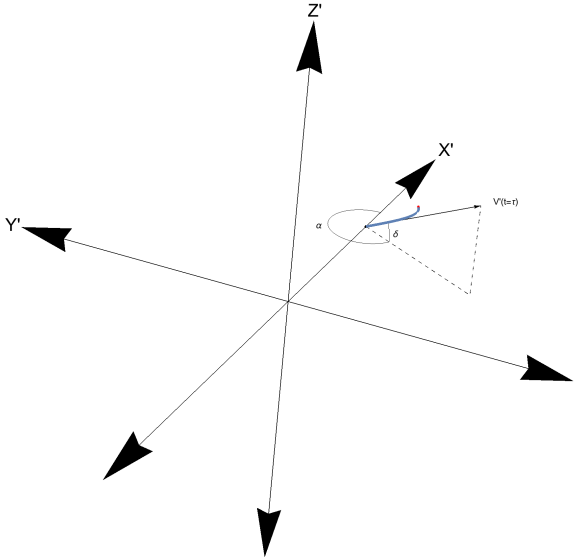}
\caption{Schematic of the wind launch angles  $\alpha$ and $\delta$ in the rotating frame. 
\label{fig:angles}}
\end{figure}

Now consider a fluid element launched  from the star's location, 
taken as a point source.  Neglecting gravitational, pressure,  and other forces, 
the fluid element  has unchanging velocity 
in the inertial frame, provided that streamlines  do not cross. 
Our neglect of gravitational 
force on the ballistically moving fluid element is most valid for wide binaries, 
where the launch speed is assumed characteristic of the depth of the gravitational well of the launching source, 
and much faster than the escape speed from the binary system. 
By assuming wind launch impulsively from a point source at time $\tau$, 
we can write a fluid trajectory equation
\begin{equation}
{\bf R}(t) = {\bf R}_*(\tau) + (t-\tau)[{\bf V}_*(\tau) + {\bf V}'_w(\alpha,\delta,\tau)],\label{eq:traj}
\end{equation}
where $t>\tau.$ The velocity 
of the fluid element  in the inertial frame is the star's velocity at the time of launch ${\bf V}_*(\tau)$, 
plus the launch velocity ${\bf V}'_w(\alpha,\delta,\tau)$ in the comoving frame of the star 
for the appropriate launch direction. 
We define launch angles in the corotating frame, with $\alpha$ being azimuthal angle counterclockwise 
relative to the $x'$ axis, and $\delta$ is the latitude above the $X'-Y'$ plane (see Figure~\ref{fig:angles}). 
The inverse transformation is used to obtain the unit vectors at the launch time:
\begin{eqnarray}
{\hat {\bf x}}'(\tau) & = &  {\hat {\bf x}} \cos\omega \tau + {\hat {\bf y}} \sin\omega \tau,\nonumber\\
{\hat {\bf y}}'(\tau) & = & - {\hat {\bf x}} \sin\omega \tau + {\hat {\bf y}} \cos\omega \tau.\label{eq:inverse}
\end{eqnarray}
If we take the initial launch of the wind in the corotating frame to be steady and isotropic, we have simply  
${\bf V}'_w(\alpha,\delta,\tau) = V_w \hat{\bf v}_\circ'.$ Here $\hat{\bf v}_\circ'$ is a unit vector pointing 
radially away from the star and given in terms of the launch angles by
\begin{equation}
\hat{\bf v}_\circ' = 
\hat{\bf x}'(\tau) \cos\delta\cos\alpha + \hat{\bf y}'(\tau)\cos\delta\sin\alpha+\hat{\bf z}(\tau)\sin\delta.\label{eq:vhat}
\end{equation}
Using equations (\ref{eq:rstar},\ref{eq:vstar},\ref{eq:traj}-\ref{eq:vhat}), 
the trajectory in the inertial frame in terms of the launch angles is given by:
\begin{eqnarray}
X & = & R_\circ \cos\omega\tau + (t-\tau) [ V_w \cos\delta\cos\beta - R_\circ\omega \sin\omega t ],\nonumber\\
Y & = & R_\circ \sin\omega\tau + (t-\tau) [ V_w \cos\delta\sin\beta + R_\circ\omega \cos\omega t],\nonumber\\
Z & = & (t-\tau) V_w \sin\delta,
\end{eqnarray}
where for brevity we have introduced $\beta \equiv \alpha+\omega\tau$.   
We nondimensionalize the equations by using length unit $R_\circ$ and time unit $R_\circ/V_w$, 
so dimensionless position is
${\bf r}={\bf R}/R_\circ$ with dimensionless components x, y, and z, 
and dimensionless velocity is ${\bf v}={\bf V}/V_w.$ We introduce the dimensionless parameter
\begin{equation}
p={{\omega R_\circ} \over {V_w}},
\end{equation} 
as the ratio of orbital speed to wind launch speed. 
Using the inverse transformation from eqs.(\ref{eq:inverse}) (applied at time $t$ rather than $\tau$),
and defining a time-like coordinate along the streamline as
\begin{equation}
s = {\frac{V_w}{R_\circ}}(t-\tau), 
\end{equation}
we obtain the (dimensionless) trajectory in the corotating frame:
\begin{eqnarray}
x' & = & \cos p s + p s \sin p s + s \cos\delta\cos(\alpha-p\,s),\nonumber\\
y' & = & -\sin p s + p s \cos p s + s \cos\delta \sin(\alpha-p\,s),\nonumber\\
z' & = & s \sin\delta.\label{eq:rvec}
\end{eqnarray}
\footnote{The coordinate $s$ does not, however, represent arc length along a streamline. 
In fact $ps$ is simply the angle in radians that the launching star 
has moved about the center of mass since launch of the fluid element at time $\tau$.} 
In spherical coordinates, 
\begin{equation}
r' = {\sqrt{1+s^2+ p^2 s^2+ 2 s \cos\delta\,(\cos\alpha  + p s \sin\alpha)}}.\label{eq:rscal}
\end{equation}

The (dimensional) velocity in the corotating frame can be obtained by taking the derivative 
of ${\bf R}'=R_\circ\,{\bf r}'$ with respect to time $t$. 
But since $\tau$ is  constant as the fluid element moves, 
we can equivalently differentiate the dimensionless position ${\bf r}'$ 
with respect to s.  This yields the primed, dimensionless  velocity components
\begin{eqnarray}
v_x' & = & \cos\delta\cos\gamma + p s [p\cos p s + \cos\delta\sin\gamma]\nonumber\\
v_y' & = & \cos\delta\sin\gamma - p s [p\sin p s + \cos\delta\cos\gamma]\nonumber\\
v_z' & = & \sin\delta,\label{eq:vvec}
\end{eqnarray}
where $\gamma \equiv \alpha-p\,s$.   
Equations (\ref{eq:rvec}) and (\ref{eq:vvec}) give a complete description of 
the wind kinematics in the corotating frame in parametric form. 
By taking the limit of no orbital motion, which corresponds to vanishing $p$, 
we can see that these equations 
yield free streaming at the wind speed in the initial launch direction. 
We also require the magnitude of the velocity, whose square is given by
\begin{equation}
v'^2 = 1  + p^2\,s \left[s (\cos^2\delta+p^2)  + 2  (\cos \alpha + p s \sin\alpha) \cos\delta \right].
\label{eq:vsquared}\end{equation}

\begin{figure}[ht!]
\includegraphics[width=8.5cm]{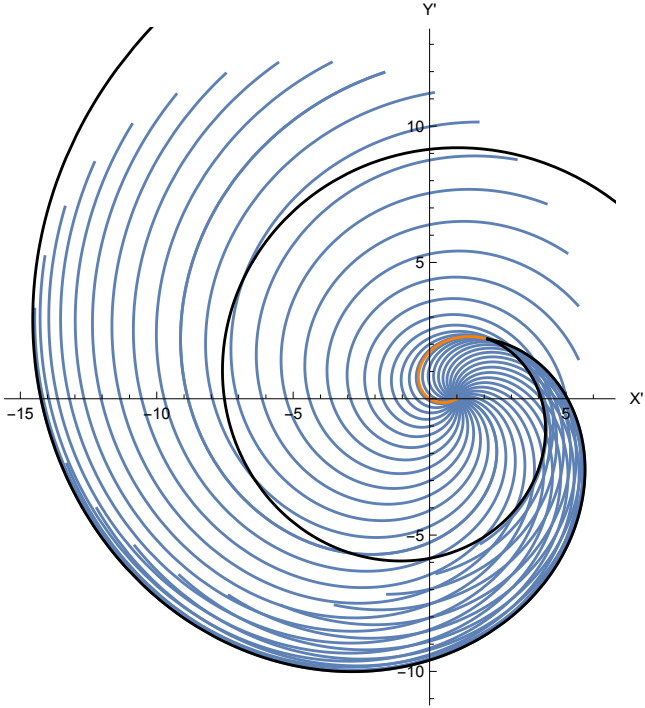}
\caption{Fluid streamlines (blue) for p=0.5 in the orbital plane ($\delta=0$).  
The black curves correspond to 
the inner and outer branches satisfying the shock condition (See \S\ref{sec:discuss}), 
and the orange curve shows the critical streamline that strikes the cusp point tangentially. 
Streamlines are drawn to a final value of $s=10$. 
\label{fig:streamlines}}
\end{figure}
\begin{figure}[ht!]
\includegraphics[width=8.5cm]{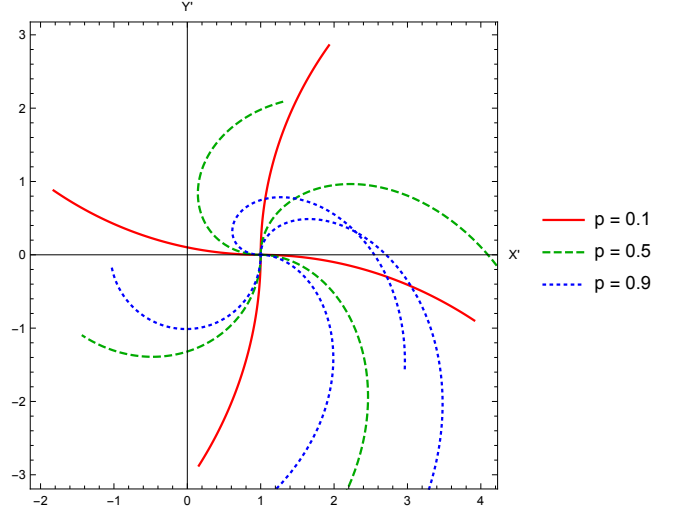}
\caption{A representation of the fluid streamlines for four launch angles $\alpha=0,\pi/2,\pi,3\pi/2$. 
For each launch angle, streamlines are 
shown  for p=0.1, 0.5, 0.9. 
\label{fig:streamlinesVariousPs}}
\end{figure}

Figure~\ref{fig:streamlines} shows the streamline shape in the rotating frame for the value $p=0.5$. 
Note the characteristic trailing spiral pattern. As will be described in \S\ref{sec:discuss}, 
streamline crossing begins at the location of the inner and outer,  solid black curves, 
depending on the $\alpha$-value of the streamline,   
and is confined to the region between them. 
Figure~\ref{fig:streamlinesVariousPs} shows the streamline shape in the rotating frame 
for four launch angles and three values of $p$. The streamlines are nearly straight for 
small $p$, and become 
increasingly distorted in a trailing spiral pattern at larger $p$ (and $s$) values. 
For sufficiently large values of  $s$ (for any $p>0$), streamlines will eventually cross, 
indicating the formation of a shock. 

\section{Derivation of the fluid density} \label{sec:density}

The behavior of the density 
$\rho$ as a function of position can be determined 
by solving the mass conservation equation in steady-state in the rotating frame. 
We introduce the density unit
\begin{equation}
\rho_\circ={\frac{\dot M_w}{4\pi R_o^2 V_w}}, \label{eq:densityunit}
\end{equation}
where $\dot M_w$ is the wind mass loss rate, so the dimensionless density is $\varrho = \rho/\rho_\circ$.
Using ${\bf v}'$, for regions where streamlines have not crossed, the mass conservation equation is
\begin{equation}
\nabla \cdot (\varrho {\bf v}') = 0.\label{eq:masscons}
\end{equation}
Our derivation will involve elements of tensor calculus 
because the most appropriate set of curvilinear coordinates will not result in an orthonormal basis. 
The reader unconcerned with the details may safely jump to equation (\ref{eq:density}), 
our result for the density.

We choose a set of coordinates $\{z^i\}=\{\alpha,\delta,s\}$, where the superscript is not 
an exponent but an index that takes values $\alpha, \delta, s.$ 
The natural (covariant) basis vectors with respect to these coordinates are given by
\begin{equation}
{\bf b}_i={{\partial {\bf r}'}\over {\partial {z^i}}},
\end{equation} 
where the position vector ${\bf r}'$ is evaluated using equations (\ref{eq:rvec}). 
The first two basis vectors are 
\begin{eqnarray}
{\bf b}_\alpha
&=&-{\hat x}' s \cos\delta \sin\gamma +{\hat y}' s \cos\delta \cos\gamma,\nonumber \\
{\bf b}_\delta
&=&- {\hat x}' s \sin\delta \cos\gamma  - {\hat y}' s \sin\delta \sin\gamma +{\hat z}' s \cos\delta,\,\label{eq:basis}
\end{eqnarray} 
while the third is  
${\bf b}_s={\bf v}',$ whose Cartesian components are given by eq.(\ref{eq:vvec}).
Because the velocity vector is entirely in the direction of ${\bf b}_s$, 
the contravariant components of the velocity vector are $v'^\alpha=v'^\delta=0$,  and $v'^s=1$. 
 
The covariant metric tensor is symmetric, with elements $g_{ij} = {\bf b}_i \cdot {\bf b}_j$ given by 
\begin{equation}
{\bf g}_ =
\left(
\begin{array}{ccc}
 s^2 \cos ^2\delta & 0 & g_{\alpha s} \\
 0 & s^2 & g_{\delta s} \\
 g_{s \alpha} 
 & g_{s \delta} 
 & v'^2 \label{eq:metrictensor}\\
\end{array}
\right),
\end{equation}
where $g_{\alpha s}=g_{s \alpha}=- p s^2 \cos \delta (p \sin \alpha+\cos \delta)$,
$g_{\delta s}=g_{s \delta}= - p^2 s^2 \cos \alpha \sin \delta$,  
and $v'^2$ is given by equation (\ref{eq:vsquared}). 
The determinant $g$ of the metric tensor 
is
\begin{equation}
g = (1+ p^2 s \cos\alpha \cos\delta)^2 s^4 \cos^2\delta. 
\end{equation}
This allows us to take a divergence using our  coordinate system.  
For a vector $\mathbf{A} = A^\alpha \mathbf{b}_\alpha + A^\delta \mathbf{b}_\delta + A^s \mathbf{b}_s$ 
the divergence is (\citet{Aris}, p.170)
\begin{equation}
\nabla \cdot \mathbf{A} = 
\frac{1}{\sqrt{g}}\left[ \frac{\partial}{\partial \alpha}\left(  \sqrt{g}\,A^\alpha \right) 
+  \frac{\partial}{\partial \delta}\left( \sqrt{g}\,A^\delta \right) 
+  \frac{\partial}{\partial s}\left(\sqrt{g} \,A^s \right) \right]
\end{equation}
If we apply this formula for divergence to the conservation of 
$\varrho\, \mathbf{v'} = \varrho\, {\bf b}_s$, according to equation (\ref{eq:masscons}),
 only the third term of the divergence 
is needed since the first two components vanish. We obtain upon integration 
\begin{equation}
\sqrt{g}\, \varrho  = f(\alpha, \delta).\label{eq:f}
\end{equation}

The function $f(\alpha, \delta)$ is determined by matching the above formula to the behavior close 
to the wind source where the flow is isotropic in the  limit of small $s$. 
Considering the mass density at a small radius 
$R_1 = R_\circ\, s_1$  centered on the source, 
\begin{equation}
\rho  = \frac{{\dot M}_w}{4 \pi R_1^2\, V_w\,}= \rho_\circ\,\frac{f(\alpha,\delta)}{\sqrt{g_1}}.
\end{equation}
For this case we have the limit $\sqrt{g_1} \rightarrow s_1^2 \cos\delta$, implying 
$f=\cos\delta$.  Equation (\ref{eq:f}) then gives the density as 
\begin{equation}
\varrho   =    {\frac{1}{s^2 \, (1+ p^2 s \cos\alpha \cos\delta)}}.\label{eq:density}
\end{equation}


\begin{figure}[ht!]
\includegraphics[width=8.5cm]{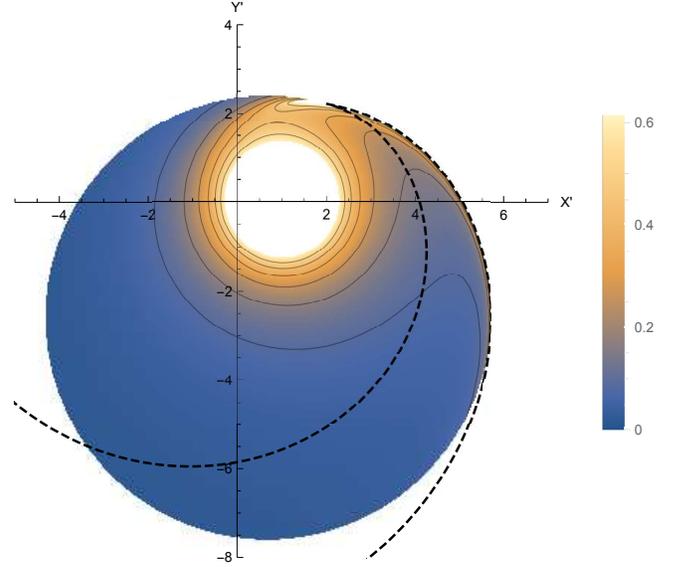}
\caption{Relative density of the rotating wind in the orbital plane, for $p=0.5$.  
The wind source is at $x=1$, $y=0$ and the center of the orbit is the origin.  
The dark curve indicates the predicted shock location. 
For a given streamline $\alpha$, the region beyond the shock is unphysical, 
so density coutours between the shocks do not describe actual density in the post-shock region. 
\label{fig:densitymap}}
\end{figure}

On more  intuitive grounds, mass conservation may be enforced in an integral sense by imagining 
a narrow streamtube,  bounded along  its length by streamlines. 
Mass may enter or leave the streamtube only at its ends: one located on a small spherical 
surface centered on the wind source, and the other at an arbitrary (but larger) constant value of s.  
Let the foot of the stream tube lie on a small sphere of radius 
$R_1 = R_0\, s_1$, and have two sides of constant $\alpha$, and $\alpha+d\alpha$, 
and two sides of constant $\delta$ and $\delta+d\delta$.  The area of the enclosed surface patch is 
$R_\circ^2\, s_1^2 \cos\delta\, d\alpha\, d\delta$. We generate a thin volume by including a thickness 
$dR = V_w\,dt = R_\circ ds.$ 
The amount of mass in this volume must equal that at the far end of 
the streamtube in a volume of matching $d\alpha$, $d\delta$, and $ds$.  
The volume element at either end of the stream tube is described by the scalar triple product
\begin{equation}
{\cal V}\,d\alpha\, d\delta\, ds 
= {\bf b}_s \cdot ( {\bf b}_\alpha \times {\bf b}_\delta )\, d\alpha\, d\delta\, ds.
\end{equation}
The cross product 
${\bf b}_\alpha \times {\bf b}_\delta \,d\alpha\,d\delta \equiv {\bf a}_s \,d\alpha\,d\delta$ 
defines a vector area element on a surface of constant $s$, where 
\begin{equation}
{\bf a}_s =  
s^2 \cos\delta \left[ {\hat x}'\,\cos\delta\cos\gamma + {\hat y}'\,\cos\delta\sin\gamma + {\hat z}'\,\sin\delta\right].
\end{equation}
Dotting this with the velocity vector ${\bf v}' = {\bf b}_s$ shows 
that the ``volume factor'' ${\cal V}$ is identical to $\sqrt{g}$, 
\begin{equation}
{\cal V}=\sqrt{g}=s^2 \cos\delta (1+p^2\,s\,\cos\alpha\cos\delta).\label{eq:rootg}
\end{equation}
Thus, in dimensional units, and using subscript `1' at the inner point, with no subscript at the outer, 
$\rho = \rho_1 \,{\cal V}_1/{\cal V} = \rho_1 \sqrt{g_1/g},$ which gives
\begin{equation}
 \rho  =  {\frac{{\dot M}_w}{4 \pi R_\circ^2\, V_w}}
{\frac{1}{s^2\,(1+p^2\,s\,\cos\alpha\cos\delta)}}.\label{eq:densityphysical}
\end{equation}
Equation (\ref{eq:densityphysical}) is the dimensional form of our previous result, eq.(\ref{eq:density}).

\section{Discussion} \label{sec:discuss}

\subsection{Streamline Crossing and Shocks}

\begin{figure}[ht!]
\includegraphics[width=8.5cm]{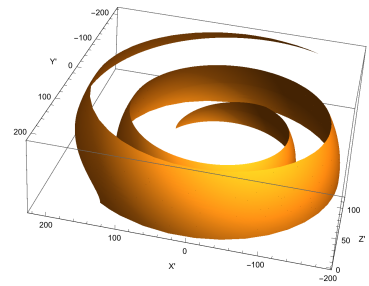}
\caption{Predicted shock as a 3-dimensional locus for $p=0.1$. 
Only launch latitudes  $0\leq \delta <\pi/6$  and azimuthal launch angle 
$\alpha$ from $2\pi/3$ to $\alpha_{cusp}$ are  shown. 
\label{fig:3D}}
\end{figure}

\begin{figure}[ht!]
\includegraphics[width=8.5cm]{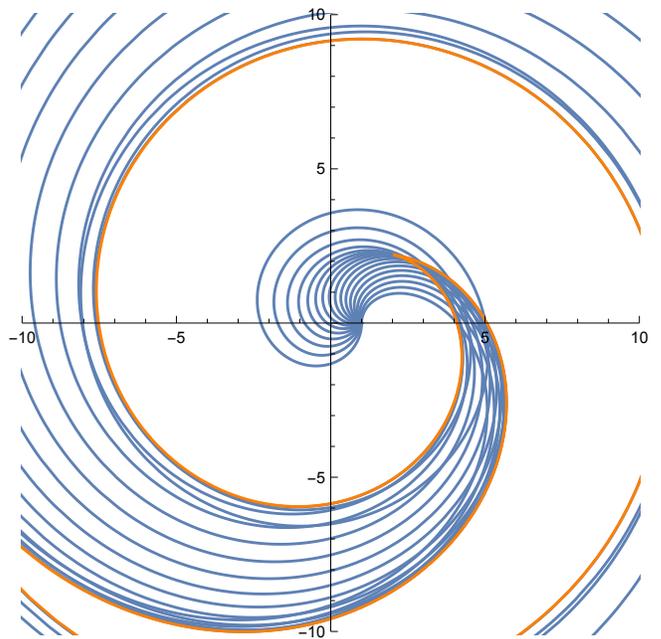}
\caption{Inner streamlines in the orbital plane for p=0.5, with $\alpha$-spacings of $\pi/16$.  
All inner streamlines eventually shock.  They may enter the shock region by crossing 
the inner or outer shock,  depending on their $\alpha$-value.  
\label{fig:innerhemisphere}}
\end{figure}

\begin{figure}[ht!]
\includegraphics[width=8.5cm]{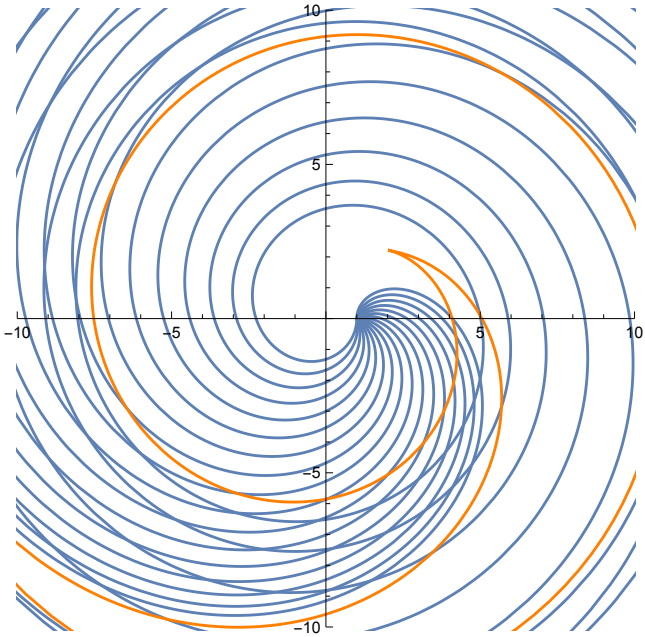}
\caption{Outer streamlines in the orbital plane for p=0.5,  with $\alpha$-spacings of $\pi/16$.  
All outer streamlines eventually cross either the  inner or outer shock. 
\label{fig:outerhemisphere}}
\end{figure}

The expression for density  contains a factor $1/s^2$ 
which for an isotropic wind from a stationary source (e.g., taking $p=0$)
results in an inverse distance-squared dependence.
But for a moving wind,  
this is more directly interpreted as an inverse square dependence on time since launch 
because for this scenario the distance  from the source is not proportional to time. 
The additional factor in parentheses 
results  in a density enhancement when $\cos\alpha<0$, and a density deficit when $\cos\alpha>0$. 
We will call these the inner (pointing toward the origin) and outer 
launch hemispheres, respectively. In fact, the density  diverges when $\sqrt{g}\,\sec\delta=0$. 
Because we aren't interested in the trivial divergence at the wind source, 
the vanishing of the factor in parentheses in equation (\ref{eq:rootg}) yields the condition 
along a given inner streamline (one for which $\cos\alpha<0$)  that 
\begin{equation}
s_{crit}=-\frac{1}{p^2 \cos\alpha\cos\delta}.\label{eq:scrit}
\end{equation}
Figure \ref{fig:3D} shows the outer branch of the predicted shock location 
for the more moderate\footnote{\citet{Soker94} estimates a range of p-values from 0.01-0.3 
for wide binaries within planetary nebulae.} value 
$p=0.1$. For outer streamlines, defined by $\cos\alpha>0$, the quantity $g$ never vanishes and the 
density remains finite.
The surface where $s=s_{crit}$ therefore  represents a shock surface, 
and the formal streamline solutions cannot be applied to collisional fluids for $s>s_{crit}$. 
The condition for local streamline crossing can also be derived by assuming an arbitrary 
set of values for ${p,\alpha,\delta, s}$ and demanding that 
two neighboring, but initially distinct fluid elements 
(the second having parameter values ${p,\alpha+d\alpha,\delta+d\delta,s+ds})$ 
have the same position using equations (\ref{eq:rvec}). 
For brevity we will not present this lengthy calculation here, but note that it yields 
$\sqrt{g}=0$, reproducing our shock condition. 
Evidently exactly half of the stellar wind, the inner launch hemisphere, ``self-shocks''  in a local sense,  
while the other half doesn't.  
The outer streamlines do, however, intersect the shock surface 
(see Figure(\ref{fig:streamlines})), 
so that eventually all streamlines 
except $\alpha=\pi/2$, $\alpha=3\pi/2$, and $\delta=\pm \pi/2$ will encounter a shock. 
The distinct trajectories of the inner and outer streamlines and their relationship 
to the inner and outer shocks 
are shown in Figures (\ref{fig:innerhemisphere}-\ref{fig:outerhemisphere}). 

An examination of Figures(\ref{fig:innerhemisphere}-\ref{fig:outerhemisphere}) 
shows that two distinct types of streamline crossings occur. The ``self-shocking'' 
involves streamlines that cross their nearest neighbors, that is, those launched with 
only a differentially distinct angle. 
These overlaps are properly termed caustics, 
and can be located with our shock criterion. 
The second type of streamline crossing we shall term ``non-local,'' 
because initially widely-spaced streamlines may cross. This includes, 
but is not limited to, all the streamline crossings involving outer streamlines. 
Outer streamlines never have self-shocking, but by crossing the caustic boundaries 
they do cross inner streamlines. Additionally, within the caustic zone inner 
streamines cross an infinite number of other inner streamlines in a continuous 
fashion. We do not have a general solution for such non-local crossings, other 
than direct solution of equations (\ref{eq:rvec}), 
which proves to be very cumbersome.  Nonetheless, 
Figures (\ref{fig:innerhemisphere}-\ref{fig:outerhemisphere}) 
show this behavior for the chosen value of p. 


\subsection{The Cusp}
The innermost point on the shock surface is a needle-like cusp of vanishing opening angle 
in the orbital plane.  The  location of the point is found 
from equations (\ref{eq:rscal},\ref{eq:scrit}) by solving for the launch angle 
resulting in minimal radius for the shock. 
The critical streamline that enters the cusp 
tangentially in the orbital plane has launch angle and $s-$value 
\begin{eqnarray}
\alpha_{cusp}& = & \pi+\sin^{-1}\,p, \\
s_{cusp} & = & \frac{1}{p^2\,\sqrt{1-p^2}}.\label{eq:scusp}
\end{eqnarray}
The location of the cusp is then found by using these values in the trajectory equation (\ref{eq:rscal}) as
\begin{equation}
r_{cusp}= \frac{1-p^2}{p^2}.\label{eq:rcusp}
\end{equation}
We note that in the limit of small $p$, equation (\ref{eq:rcusp}) is consistent 
with the results of \citet{Raga11} and \citet{Canto99}, 
who assumed a purely radial wind of time-dependent speed, launched from the center of mass, 
obtaining in dimensional units $R_{cusp}=V_w^2/R_\circ\omega^2$. Our result generalizes 
this with an exact treatment of centrifugal and coriolis effects. 
\citet{KimTaam2012b} also give an empirical formula predicting a smaller standoff 
location based upon numerical simulations, although these simulations also incorporated gravitational 
perturbations. 
The discrepancy may be associated with the difficulty of determining the innermost location along 
the shock, because 
the convergence of streamlines prior to reaching the cusp results in a density enhancement 
which may have been 
interpreted as a cusp in their simulations. 
For $p>1$, no cusp occurs, and there will be regions unreached by the flow due to a centrifugal barrier.  
In particular, note that the same streamline that intersects 
the cusp point also goes through the origin. For $p>1$, the flow does not reach the origin, 
with implications for identical colliding winds: no stagnation point region would be possible. 
Because our neglect of gravitational forces was justified 
by application to wide binaries only (small $p$), we will not discuss the case $p>1$. 


\subsection{The Shocked Region}
Our streamline analysis does not provide a description of the region between the two shock surfaces. 
However, an asymptotic, ``tail''  solution 
for the shock may be developed for the outer part of the shock spiral by noting that the 
azimuthal range $\pi/2<\alpha<3\pi/2$ is mapped 
to an infinite range of azimuthal angle  
in the outer region,  so that most of the spiral corresponds to a 
small $\epsilon>0$ region of either $\alpha=\pi/2+\epsilon$ or $\alpha=3\pi/2-\epsilon$. 
Physically, this means most of the wind mass 
has reached the shock before many windings occur. 
Either limit yields $s_{crit} \approx 1/p^2\epsilon\,\cos\delta \equiv \zeta/p$ by direct substitution 
into equation (\ref{eq:scrit}) and expansion to leading order. 
Substitution  into equations (\ref{eq:rvec}) gives  outer solution from $3\pi/2-\epsilon$, 
and inner solution from $\pi/2+\epsilon$, as 
\begin{eqnarray}
 {\bf r}'_{tail} & \approx & (\zeta/p) [({\hat {\bf x}}'\,(\pm\,\cos\delta\sin\,\zeta+p\,\sin\,\zeta)\nonumber \\
& + & {\hat {\bf y}}'\,(\pm\,\cos\delta\cos\,\zeta+p\,\cos\,\zeta)\nonumber \\
& + & {\hat {\bf z}}'\,\sin\delta],\label{eq:xyztail}
\end{eqnarray}
which yields radii for the two tails 
\begin{eqnarray}
r_{tail} & = & (\zeta/p) \sqrt{1\pm 2p\cos\delta+p^2}.\label{eq:zetatwotails}
\end{eqnarray}
Both tail solutions take the form of an Archimedean spiral
where $\zeta$ plays the role of azimuthal angle. 

The  radial spacing for a given curve is found by adding argument $2\pi$ to $\zeta$, 
which gives 
\begin{equation}
\Delta\,R=2\pi\,R_\circ\,\sqrt{1\pm 2p\cos\delta+p^2}/p.
\end{equation}
This radial step 
is a generalization of the expression given by \citet{MauronHuggins,Raga11} 
and agrees with both for vanishing $p$. 
On the other hand, the radial spacing (fixed $\zeta$ and $\delta$) between the two shock curves 
provides an estimate of the thickness of the shell of shocked gas. 
When evaluated for small $p$,  this radial thickness is $\Delta R \approx R_\circ \zeta \cos\delta$, 
and using $R\approx R_\circ\,\zeta/p$, the fractional thickness is
$\Delta R/R \approx p\cos\delta.$ 
Because the radial step for the inner shock is smaller than that for the outer shock, 
and these steps are approximately constant, 
the mis-match implies that eventually the outer shock for, say, 
n orbits, should cross the inner shock for n+1 orbits, 
resulting in a pinching-off of all unshocked streamlines. 
The intersection of two shocks may produce filamentary or spoke-like features, although for large enough intersection angle a Mach stem could result  (e.g.\citet{2016ApJ...823..148H}). 
Given the complex  spiral geometry, 
we cannot anticipate the details of such Mach structures here, and leave them 
for further study that could be done with 3-d hydrodynamics simulations.



\subsection{Summary}

We have considered the flow structure of an initially isotropic wind from a point source in a circular orbit. 
Assuming ballistic flow, the pre-shock particle trajectories 
are straight lines in the inertial frame of the center of mass, 
and trailing spirals in the frame rotating with the source. 
In the corotating frame, the flow is steady-state and trajectories are 
initially non-intersecting streamlines. By solving the mass conservation condition in a curvilinear coordinate 
system based upon these streamlines, we obtain an exact solution for the density structure. This density has a 
latitude and initial launch angle dependence. Interestingly, the usual $1/r^2$ dependence is modified 
as an inverse square dependence on time of flight, modified by a geometric factor. 

The flow is separated into two behaviors. The inner launch hemisphere generates a converging flow so that
streamlines eventually intersect with their neighbors,
resulting in a shock. The outer hemisphere of 
launch has non-intersecting, diverging streamlines. However, outer streamlines do eventually run into the shock region. 
A local criterion for streamline crossing gives the region 
beyond which the description is incomplete. 
A full description of the post-shock flow as a spiral shell is beyond the scope  of this study. 


The generic nature of our model and the ubiquity  of winds in widely-separated binaries 
imply that our solution should  be applicable to modeling a wide variety of observed sources 
such as spiral nebulae of AGB stars, 
structures within planetary nebulae, and colliding winds in binary 
systems.

\acknowledgments

H.H. acknowledges financial support from the Lee L. Davenport  Fellowship.

\end{document}